\documentclass{article}

     \usepackage[preprint]{neurips_2018}

\usepackage[utf8]{inputenc} 
\usepackage[T1]{fontenc}    
\usepackage{hyperref}       
\usepackage{url}            
\usepackage{booktabs}       
\usepackage{amsfonts}       
\usepackage{nicefrac}       
\usepackage{microtype}      
\usepackage{color}
\usepackage{amsmath}
\usepackage{amsfonts}
\usepackage{cite}

\title{HAXMLNet: Hierarchical Attention Network for Extreme Multi-Label Text Classification}

\author{
	Ronghui You \\
	School of Computer Science \\
	Fudan University \\
	\texttt{rhyou18@fudan.edu.cn}
	\And
	Zihan Zhang \\
	School of Computer Science \\
	Fudan University \\
	\texttt{17210240027@fudan.edu.cn}
	\And
	Suyang Dai \\
	School of Computer Science \\
	Fudan University \\
	\texttt{16210240005@fudan.edu.cn}
	\And
	Shanfeng Zhu \\
	School of Computer Science \\
	Fudan University \\
	\texttt{zhusf@fudan.edu.cn}
}

\begin{document}

\maketitle

\begin{abstract}
Extreme multi-label text classification (XMTC) addresses the problem of tagging each text
with the most relevant labels from an extreme-scale label set. Traditional methods
use bag-of-words (BOW) representations without context information as their 
features. The state-ot-the-art deep learning-based method, AttentionXML, which 
uses a recurrent neural network (RNN) and the multi-label attention, can hardly
deal with
extreme-scale (hundreds of thousands labels) problem. To address this, we propose our HAXMLNet,
which uses an efficient and effective hierarchical structure with the multi-label 
attention. Experimental results show that HAXMLNet reaches a competitive performance with other state-of-the-art methods.
\end{abstract}

\section{Introduction}

Extreme multi-label text classification (XMTC) addresses the problem of tagging 
each document with the most relevant labels from an extreme-scale label set. For 
capturing context information, long-term dependency and the most relevant part 
for each label, the state-of-the-art deep learning-based method, 
AttentionXML\citep{you2018attentionxml} is proposed. AttentionXML used a recurrent 
neural network (RNN) with a multi-label attention. For given context inputs 
$\mathbf{h}_i\in\mathbb{R}^n$, multi-label attention gets output as follows:
\begin{equation}
\begin{split}
\alpha_{ij} &= \frac{e^{\mathbf{h}_i\mathbf{w}_j}}{\sum_{t=1}^T e^{\mathbf{h}_{t}\mathbf{w}_j}}, \\
\mathbf{m}_j &= \sum_{i=1}^T \alpha_{ij}\mathbf{h}_i, \\
\end{split}
\end{equation}
where $\alpha_{ij}$ is the normalized coefficient of $\mathbf{h}_i$ and
$\mathbf{w}_j\in\mathbb{R}^{n}$ is the so-called attention parameters.
However, AttentionXML cannot handle extreme-scale datasets.

In this paper, we proposed HAXMLNet, keeping label-wise attention and using a
 \textbf{probabilistic label tree(PLT)}\citep{jasinska2016extreme} for solving 
extreme-scale datasets. Experimental results show that HAXMLNet reaches
a competitive performance on two extreme-scale datasets with other state-of-the-art methods.

\section{Method: HAXMLNet}

Directly using AttentionXML\citep{you2018attentionxml} can hardly deal with solve extreme-scale 
multi-label classification (with hundreds of thousands labels) due to its model scale 
(can't put in GPU memory) and computational complexity (slow training and prediction). 
For solving classification with extreme-scale labels, we proposed an efficient and 
effective method, HAXMLNet, which also uses label-wise attention with a 
\textbf{PLT} to reduce the model scale and computational complexity during training and 
prediction as follows:
\begin{enumerate}
\item 
We partitioned all labels into $g$ groups. The difference of size among each group is 
less than one. Each label $j$ belongs to only one group $G(j)$ as its father node, which 
is called group label for label $j$. We add a root node as father of all group label and 
construct a \textbf{PLT} with height of 3 (including root node).
\item
For each training sample, we generate its group labels, which are all group labels of 
its raw labels. Using group labels as target labels, it's still a multi-label 
classification problem, but notice that the size of group labels are $|\frac{L}{g}|$ 
times smaller than the original raw labels. We train an AttentionXML called HAXMLNet-G with these 
group labels.
\item
We use all labels in groups of each training sample as its candidate labels. Obviously,
candidate labels of each sample includes its all raw (positive) labels and some negative labels. We train another AttentionXML named HAXMLNet-L by using these 
candidate labels only. During training process, we calculate attention and loss scores 
only with these candidate labels. The loss function of HAXMLNet-L 
is given as follows:
\begin{equation}
\small
J(\theta) = -\frac{1}{N|C(i)|}\sum_{i=1}^N\sum_{j\in C(i)} y_{ij}log(\hat{y}_{ij})+(1-y_{ij})log(1-\hat{y}_{ij}),
\end{equation}
where $C(i)$ is the candidate labels of $i_{th}$ sample, $y_{ij}$ and $\hat{y}_{ij}$ 
are the ground truth and predicted probability, respectively. Note that $\hat{y}_{ij}$
is a conditional probability because we used a $PLT$ and we already know that $G(j)$ is 
a truth group label of sample $i$. The expectation number of candidate labels for each 
sample is the average number of group labels per sample (no more than the average number 
of raw labels) timing the group size, which is much smaller than the number of all labels. 
We also use a number $\textbf{c=1000}$, if candidate number of this sample is larger than 
$c$, we only keep $c$ candidate labels randomly.
\item
During prediction, for a given 
sample, the predicted score $\hat{y}_j$ for $j_{th}$ label based on probability chain rule 
is as follows:
\begin{equation}
\hat{y_j} = \hat{y}^{group}_{G(j)} \times\hat{y}^{label}_j
\end{equation}
where $\hat{y}^{group}_{G(j)}$ is the predicted score for the group $G(j)$ by 
AttentionXML-G, and $\hat{y}^{label}_j$ is the predicted score for the $j_{th}$
label by AttentionXML-L. For the efficiency of our model, we only predicted
the scores for labels in top \textbf{\textbf{k=10}} groups.
\item 
If complexity of AttentionXML-G is still too large, we can use the above algorithm on this
multi-label classification problem recursively. 
\end{enumerate}

Here we train two AttentionXML models (HAXMLNet-G and HAXMLNet-L), but notice that
the complexity of both HAXMLNet-G and HAXMLNet-L are much smaller 
than AttentionXML against all raw labels directly. \textbf{If the scale of HAXMLNet-G
is still too large, we use this algorithm recursively because training HAXMLNet-G is also
a multi-label classification task.}

Inspired by Parabel \citep{prabhu2018parabel}, we use a similar top-down hierarchical 
clustering to constructing a PLT rather than random construction. This clustering approach 
divided a label clustering into two clusters with a balanced k-means and repeated 
recursively until the size of all clusters smaller than a given leaf size. Different from 
Parabel, we only keep the leaves of this tree as its clustering result. Now we can solve 
extreme-scale datasets by label-wise attention efficiently.

\section{Experiments}

\subsection{Dataset}

We used two extreme-scale multi-label datasets benchmarks: Amazon-670K and Wiki-500K 
(Table \ref{tab:dataset}). All these datasets are downloaded from \textbf{XMLRepository}
\footnote{\url{http://manikvarma.org/downloads/XC/XMLRepository.html}}.

\begin{table*}[t]
   \vspace{-2mm}
	\centering
	\caption{Datasets we used in our experiments.}
	\label{tab:dataset}
	\begin{tabular}{@{}crrrrrrrrrrrrrrrrrr@{}}
	\hline
	Dataset & $N_{train}$ & $N_{test}$ & $D$ & $L$ & $\overline{L}$ & $\hat{L}$ & $\overline{W}_{train}$ & $\overline{W}_{test}$ \\
	\hline
	Amazon-670K & 490,449 & 153,025 & 135,909 & 670,091 & 5.45 & 3.99 & 247.33 & 241.22 \\
	Wiki-500K & 1,779,881 & 76,9421 & 2,381,304 & 501,008 & 4.75 & 16.86 & 808.66 & 808.56 \\
	\hline	
	\end{tabular}

 $N_{train}$: \#training instances,
 $N_{test}$: \#test instances, $D$: \#features, $L$: \#labels,
 $\overline{L}$: average \#labels per instance, $\hat{L}$: the average
 \#instances per label, $\overline{W}_{train}$: the average \#words
 per training instance and $\overline{W}_{test}$: the average \#words
	per test instance. The partition of training and test is from the data source.
\vspace{-3mm}
\end{table*}

\subsection{Evaluation Metric}

We chose $P@k$ (Precision at $k$) and $nDCG@k$ (normalized Discounted
Cumulative Gain at $k$) as our evaluation metrics for performance 
comparison, since both $P@k$ and $nDCG@k$ are widely used for evaluation 
methods for multi-label classification problems. $P@k$ is defined as follows:
\begin{equation}
P@k = \frac{1}{k} \sum_{l=1}^k \mathbf{y}_{rank(l)}
\end{equation}
where $\mathbf{y} \in \{0, 1\}^L$ is the true binary vector, and
$rank(l)$ is the index of the $l$-th highest predicted label.
$nDCG@k$ is defined as follows:
\begin{equation}
\begin{split}
DCG@k &= \sum^{k}_{l=1} \frac{\mathbf{y}_{rank(l)}}{log(l+1)} \\
iDCG@k &= \sum^{min(k, ||\mathbf{y}||_0)}_{l=1} \frac{1}{log(l+1)} \\
nDCG@k &= \frac{DCG@k}{iDCG@k}
\end{split}
\end{equation}
$nDCG@k$ is a metric for ranking, meaning that the order of top $k$
prediction was considered in $nDCG@k$ but not in $P@k$.
Note that P@1 and nDCG@1 are the same. 

\subsection{Performance}

Table \ref{tab:per:amzon} and Table \ref{tab:per:wiki} show performance
comparison of HAXMLNet and several state-of-the-art methods on Amazon-670K 
and Wiki-500K respectively. Note that Parabel here used 3 trees while 
HAXMLNet only used 1 tree. On Wiki-500K, HAXMLNet reaches the best 
performance on all experimental settings. Because of high sparsity of labels
in Amazon-670K (the most frequent label only has less than 1,900 positive 
samples), performance of HAXMLNet is worse than DiSMEC and Parabel, but still  
have a competitive performance (better than PfastreXML). 

\begin{table*}
\centering
\caption{Performance comparison of HAXMLNet and other competing methods on Amazon-670K.}
\label{tab:per:amzon}
\begin{tabular}{@{}ccccccccccccccccccccccc@{}}
\hline
Methods & Prec@1 & Prec@3 & Prec@5 & nDCG@1 & nDCG@3 & nDCG@5 \\
\hline
AnnexML\citep{tagami2017annexml} & 42.08 & 36.66 & 32.74 & 42.08 & 38.81 & 36.78 \\ 
PfastreXML\citep{prabhu2018extreme} & 36.84 & 34.24 & 32.09 & 36.84 & 36.02 & 35.43 \\
DiSMEC\citep{babbar2017dismec} & \textbf{45.40} & \textbf{40.42} & \textbf{36.97} & \textbf{45.40} & \textbf{42.83} & \textbf{41.37} \\
Parabel\citep{prabhu2018parabel} & 44.92 & 39.77 & 35.98 & 44.92 & 42.11 & 40.34 \\
XML-CNN\citep{liu2017deep} & 33.41 & 30.00 & 27.41 & 33.41 & 31.78 & 30.67 \\
HAXMLNet & 41.09 & 36.39 & 32.65 & 41.09 & 38.49 & 36.64 \\
\hline
\end{tabular}
\end{table*}

\begin{table*}
\centering
\caption{Performance comparison of HAXMLNet and other competing methods on Wiki-500K.}
\label{tab:per:wiki}
\begin{tabular}{@{}ccccccccccccccccccccccc@{}}
\hline
Methods & Prec@1 & Prec@3 & Prec@5 & nDCG@1 & nDCG@3 & nDCG@5 \\
\hline
AnnexML\citep{tagami2017annexml} & 64.22 & 43.15 & 32.79 & 64.22 & 54.32 & 52.25 \\ 
PfastreXML\citep{prabhu2018extreme} & 56.25 & 37.32 & 28.16 & 56.25 & 47.14 & 45.05 \\
DiSMEC\citep{babbar2017dismec} & 64.77 & 45.14 & 35.08 & 64.77 & 55.85 & 54.17 \\
Parabel\citep{prabhu2018parabel} & 68.79 & 49.57 & 38.64 & 68.79 & 60.54 & 58.60 \\
HAXMLNet & \textbf{70.44} & \textbf{52.42} & \textbf{41.12} & \textbf{70.44} & \textbf{62.91} & \textbf{60.80} \\
\hline
\end{tabular}
\end{table*}

\section{Conclusion}
In this paper, we propose HAXMLNet, a hierarchical label-wise attention network,
which can solve extreme multi-label text classification efficiently and effectively.
HAXMLNet uses a hierarchical structure with multi-label attention to reduce the 
scale and complexity of the state of the art method AttentionXML, and make a 
competitive performance with other state of the art methods like Parabel
\citep{prabhu2018parabel} and DiSMEC\citep{babbar2017dismec}. Future work will 
focus on the new algorithms to improve the performance on extreme-scale multi-label 
classification.

\bibliographystyle{abbrv}
\bibliography{HAXMLNet}

\end{document}